The paper will be presented at the ***2027 21th European Conference on Antennas and Propagation (EuCAP)***, Düsseldorf, Germany, April 18–23, 2027

# Static Multipath Suppression in Power Delay Profile of Time-Varying Channel at 28 GHz

Jarosław Wojtuń[1], Bartosz Czaja[1], Tomasz Grasza[1], Jan M. Kelner[1], Cezary Ziółkowski[1], Radek Zavorka[2], Josef Vychodil[2], Tomas Mikulasek[2], Jiri Blumenstein[2], Ales Prokes[2], and Aniruddha Chandra[3]

[1] Institute of Communications Systems, Faculty of Electronics, Military University of Technology, Warsaw, Poland, *jaroslaw.wojtun@wat.edu.pl*

[2] Department of Radio Electronics, Brno University of Technology, Brno, Czech Republic

[3] ECE Department, National Institute of Technology, Durgapur, India

***Abstract*—This paper investigates static multipath suppression in a time-varying radio channel at 28 GHz using wideband measurements conducted in a dynamic road environment. Two temporal differencing filters and three Doppler-domain notch filtering configurations are compared in terms of their effects on power delay profiles (PDPs) and root mean square (RMS) delay spread. The results demonstrate that all investigated methods effectively suppress stationary multipath components, improving the visibility of vehicle-related reflections. However, differences in filter frequency responses influence the relative powers of dynamic multipath components and the resulting RMS delay spread distributions. The findings highlight the importance of static multipath suppression and filter selection in characterizing time-varying propagation channels.**

## I. Introduction

The characterization of time-varying radio propagation channels is essential for understanding the influence of moving objects on wideband wireless systems. At millimeter-wave (mmWave) frequencies, high temporal resolution enables the identification of individual multipath components associated with stationary and moving scatterers. The channel impulse response (CIR), power delay profile (PDP), and root mean square (RMS) delay spread are commonly used to describe the resulting temporal dispersion. However, in environments containing both stationary infrastructure and moving objects, strong persistent multipath components may dominate the received signal and obscure weaker time-varying contributions.

Wideband channel sounding and the estimation of temporal dispersion parameters have been extensively discussed by Salous [1]. Gentile *et al.* [2] investigated the influence of channel-sounder characteristics and measurement procedures on the reliability of propagation measurements, while Quimby *et al.* [3] examined channel-sounder performance verification using PDPs and derived channel parameters, including RMS delay spread. These studies emphasize the importance of reliable channel measurements and signal processing. In dynamic propagation scenarios, an additional challenge arises from separating persistent multipath contributions from those associated with moving scatterers. Temporal differencing and Doppler-domain filtering provide two possible approaches to suppressing approximately stationary components, although their distinct frequency responses may also affect the remaining dynamic multipath structure.

In wideband channel sounding measurements, the CIR can be estimated by cross-correlating the received signal with a known sounding waveform. For discrete-time signals, the estimated CIR is expressed as

$$h(\tau_k) = \sum_{n=0}^{N-1} r(t_n) s^*(t_n - \tau_k), \qquad (1)$$

where $r(t_n)$ denotes the received signal, $s(t_n - \tau_k)$ is the time-shifted sounding waveform, and $(\cdot)^*$ denotes complex conjugation. The sampling instants and discrete delays are defined as $t_n = n/f_s$, and $\tau_k = k/f_s$, respectively, where $f_s$ is the sampling frequency and $N$ is the number of received samples.

When successive CIR estimates are acquired over time, stationary multipath contributions typically remain concentrated around constant delays, whereas reflections from moving objects exhibit time-dependent delays and amplitudes. Suppressing the persistent components may therefore improve the visibility of dynamic propagation paths. However, the applied filtering operations can also modify the amplitudes of moving-scatterer contributions, potentially affecting the estimated PDP and RMS delay spread.

This paper investigates static multipath suppression in a time-varying mmWave channel at 28 GHz using experimental measurements conducted in a dynamic road environment. Two temporal differencing filters and three Doppler-domain notch filtering configurations are compared using the same recorded channel data. Their effects are evaluated through the resulting PDPs, the temporal evolution of RMS delay spread, and its empirical cumulative distribution functions. The objective is to assess the effectiveness of the investigated methods in suppressing stationary multipath components and to determine how the choice of filtering approach influences temporal dispersion characterization.

The remainder of this paper is organized as follows. Section II describes the measurement setup and propagation

This research was funded in part by the National Science Center (NCN), Poland, grant no. 2021/43/I/ST7/03294 (MubaMilWave).

scenario. Section III introduces the investigated static multipath suppression methods. Section IV presents the experimental results and comparative analysis, while Section V summarizes the main conclusions.

## II. Measurement Campaign

### *A. Measurement Setup*

The measurements were performed using a wideband mmWave channel sounder comprising a Keysight M9384B VXG vector signal generator (Tx) and a Keysight N9032B PXA signal analyzer (Rx), as illustrated in Fig. 1. Both instruments operate up to 44 GHz and support instantaneous bandwidths of 1 GHz. The Tx chain included an Eravant SBP-2034434526-2F2F-S1 Ka-band power amplifier (45 dB nominal gain, 20–44 GHz), while the Rx chain employed an Eravant SBL-2434434025-2F2F-E3 low-noise amplifier (40 dB nominal gain, 24.25–43.5 GHz). Low-loss coaxial cables of 5 m and 0.5 m were used between the instruments, amplifiers, and antennas on both sides. Time and frequency synchronization was provided by a Stanford Research Systems FS740 GPS-disciplined reference.

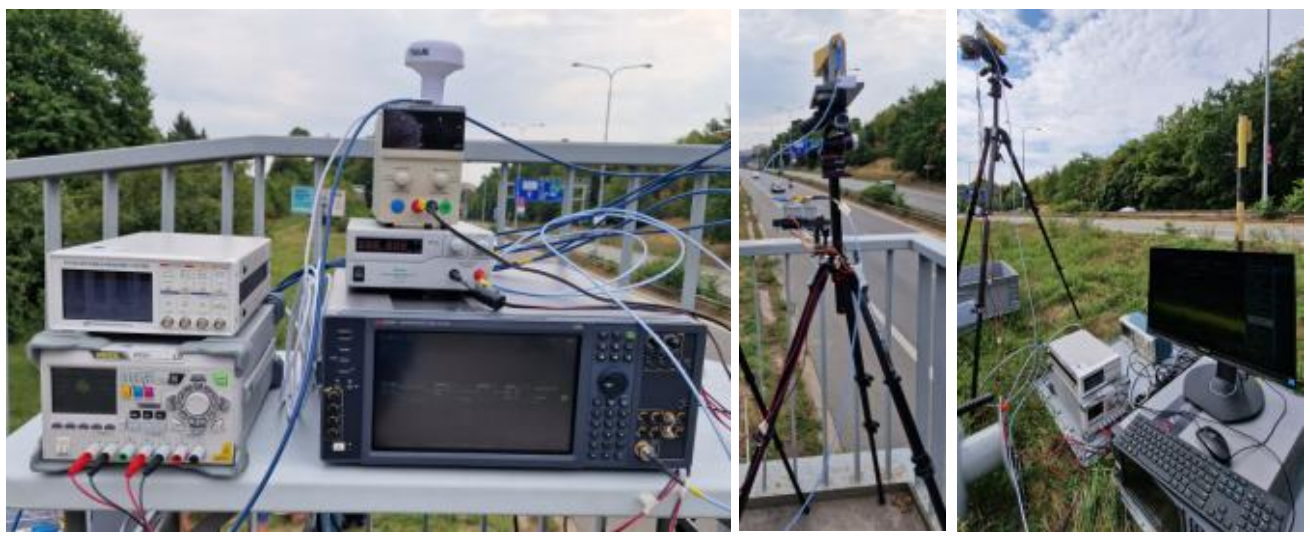

(a) (b) (c)

Fig. 1. Measurement testbed: (a) TX subsystem, (b) TX antenna, (c) RX subsystem.

Identical Eravant SAR-2507-28-S2 horn antennas were employed at both ends, providing a nominal gain of 25 dBi over 26.5–40 GHz and half-power beamwidths of approximately 7° and 9° in elevation and azimuth, respectively.

Channel sounding employed an 8 µs frequency-modulated continuous-wave (FMCW) chirp with a bandwidth of 1 GHz. Both instruments operated at 1.2 GSa/s. A Tektronix AFG3021 function generator triggered 16 µs Rx acquisition windows every 25 ms, corresponding to a 40 Hz observation rate. Each window contained at least one complete chirp. The 30 s measurement yielded 1,200 channel observations at a carrier frequency of 28 GHz.

### *B. Measurement Scenario*

The measurements were conducted on August 26, 2026, along Hradecká Street (640/E461), near the Brno University of Technology campus, Czech Republic (see Fig. 2). The Tx antenna was installed approximately 5 m above road level on a pedestrian footbridge connecting Technická and Purkyňova Streets, while the Rx antenna was positioned near the roadside at a height of approximately 1.5 m. Both antennas were directed toward approaching vehicles along the nearest traffic lane. The relatively small Tx–Rx separation compared with the distances to the observed vehicles resulted in a quasi-monostatic geometry. The direct Tx–Rx path remained outside the main antenna beams, and vehicle-related propagation components were observed under non-line-of-sight conditions.

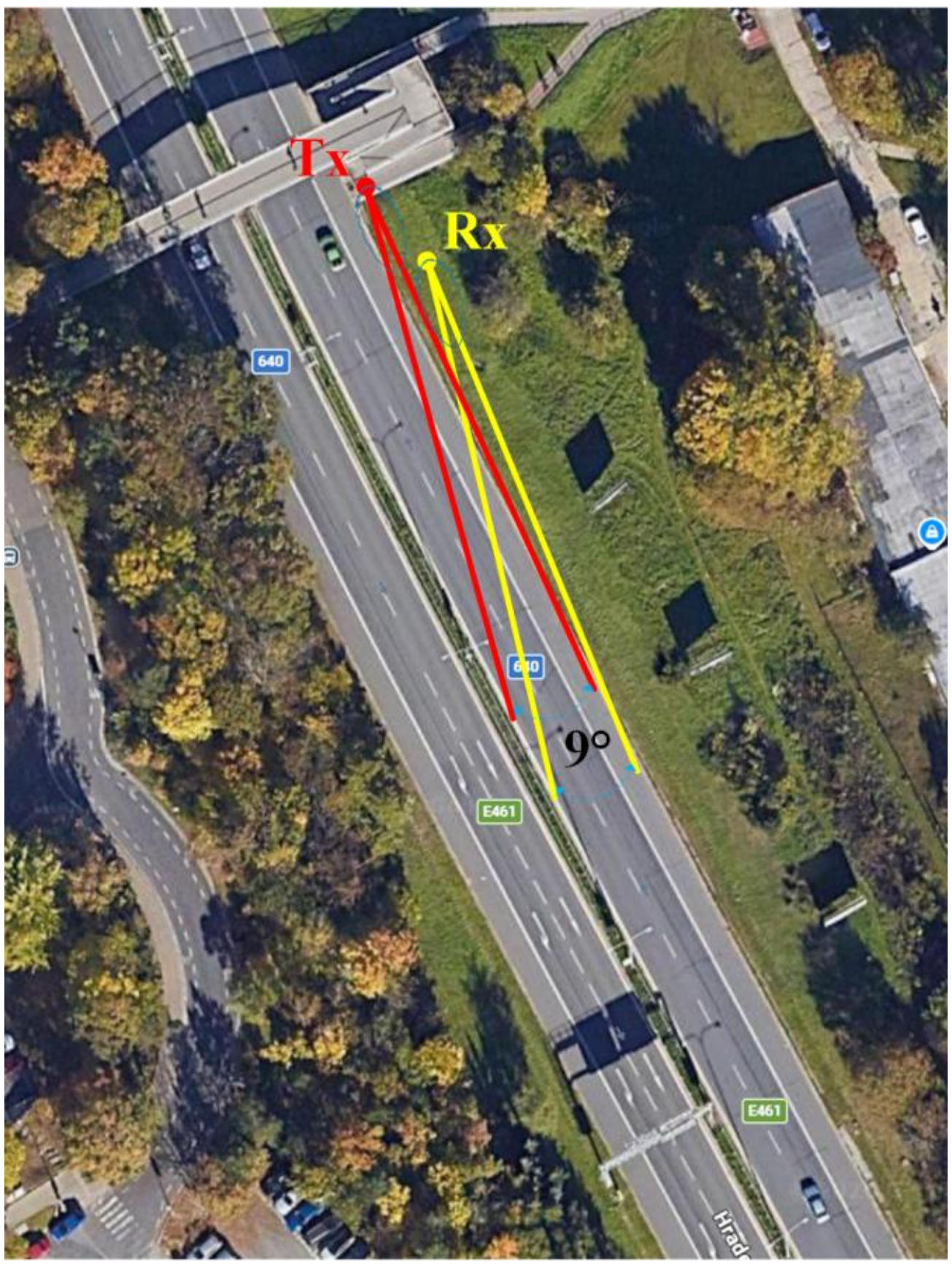

Fig. 2. Aerial view of measurement scenario along Hradecká Street near BUT campus, showing Tx and Rx locations and monitored traffic direction (background map: Google Maps).

The investigated environment was relatively open but contained several stationary objects capable of producing persistent multipath contributions. These included street lighting poles, traffic signs, overhead road information boards, and metallic safety barriers separating opposite traffic directions. Reflections and scattering from these objects contributed to the relatively stable part of the received channel response. In contrast, vehicles entering, traversing, and leaving the antenna observation region generated time-dependent propagation components whose delays and received powers varied with vehicle position and movement.

The resulting channel response therefore comprised a combination of relatively persistent multipath components associated with stationary objects and transient contributions caused by moving vehicles. Their relative strengths and delay characteristics changed throughout the observation period, depending on the instantaneous traffic conditions. This measurement scenario provides a suitable basis for investigating the suppression of static multipath components in the PDP and assessing their influence on the characterization of time-varying channel dispersion [4].

A more detailed description of the measurement scenario and setup is presented in [4], [5].

## III. Static Multipath Suppression Methods

In dynamic propagation environments, the estimated CIR contains multipath contributions from both stationary and moving scatterers. Static components appear as approximately constant-delay paths and may mask weaker reflections from passing vehicles. To suppress these components, two temporal differencing filters and a Doppler-domain notch filter were considered.

### *A. Temporal Differencing*

Temporal differencing suppresses multipath components that remain approximately constant between consecutive channel observations. Let $h_m(\tau_k)$ denote the estimated complex CIR at delay $\tau_k$ and observation index $m$. First-order differencing is implemented using a finite impulse response (FIR) filter with coefficients [1, –1]:

$$h_m^{(1)}(\tau_k) = h_m(\tau_k) - h_{m-1}(\tau_k) \tag{2}$$

Second-order differencing uses the FIR coefficients [1, –2,1]:

$$h_m^{(2)}(\tau_k) = h_m(\tau_k) - 2h_{m-1}(\tau_k) + h_{m-2}(\tau_k) \tag{3}$$

Both filters eliminate ideally constant components. The second-order filter additionally suppresses components varying approximately linearly between successive observations. However, temporal differencing also modifies the amplitudes of dynamic multipath components, depending on their rate of variation. The resulting PDPs are calculated as $P_m^{(i)}(\tau_k) = \left|h_m^{(i)}(\tau_k)\right|^2$, where ($i$=1,2). The first one or two observations, respectively, are excluded because the required preceding samples are unavailable.

Both differencing filters exhibit a zero at zero Doppler frequency and therefore suppress ideally stationary multipath components. However, their frequency responses differ: second-order differencing provides stronger attenuation of slowly varying components, while both filters modify the amplitudes of dynamic multipath contributions in a Doppler-dependent manner. Consequently, differences between the resulting PDPs reflect not only static-component suppression but also the distinct frequency-selective characteristics of the applied filters.

### *B. Doppler-Domain Filtering*

Doppler-domain filtering exploits differences in the temporal variation of static and dynamic multipath components. For each delay bin, the discrete Fourier transform (DFT) of the complex CIR is calculated along the observation-time axis. Static components are expected to concentrate around zero Doppler frequency, while moving scatterers generally produce nonzero Doppler frequencies.

A rectangular notch filter centered at zero Doppler frequency is applied as

$$\widetilde{H}(f_D, \tau_k) = \begin{cases} 0, & |f_D| \le f_N, \\ H(f_D, \tau_k), & |f_D| > f_N, \end{cases} \tag{4}$$

where $H(f_D, \tau_k)$ represents the Doppler spectrum of the CIR at delay $\tau_k$, and $f_N$ denotes the notch half-width. The filtered CIR is reconstructed using the inverse DFT, and the corresponding PDP is obtained from its squared magnitude.

Three notch half-widths, $f_N = \{0.9, 1.4, 1.9\}$ Hz, were investigated to assess the influence of the suppressed Doppler-frequency range on the resulting PDP and RMS delay spread estimates. Increasing the notch width allows progressively wider suppression of stationary and slowly varying multipath components.

The Doppler-domain processing was performed over the complete 30 s observation interval, comprising 1,200 channel snapshots acquired at a rate of 40 Hz. The resulting Doppler-frequency resolution was approximately 0.033 Hz, allowing narrowband suppression around zero Doppler frequency.

### *C. Summary*

All filtering methods were applied to the complex CIR prior to PDP calculation. Since the filters exhibit different amplitude responses, the resulting PDPs may differ in both the suppression of stationary components and the relative power distribution of dynamic multipath contributions. These effects should be considered when comparing the corresponding RMS delay spread estimates.

## IV. Results

The effects of static multipath suppression were evaluated by comparing the original and filtered PDPs over the delay interval of 0–1 μs. Fig. 3 presents the unfiltered PDP, revealing persistent horizontal traces associated with stationary scatterers and inclined trajectories corresponding to moving vehicles. Downward-sloping traces primarily represent approaching vehicles, whereas upward-sloping traces are associated with receding vehicles. Their slopes are related to the vehicles' radial velocities, while their intensities depend on the scattering characteristics and propagation geometry.

All investigated filtering methods substantially attenuate stationary multipath contributions, including a strong component at approximately 0.7 μs, likely associated with an overhead road information board. The PDPs obtained using first- and second-order FIR filtering exhibit similar structures, as do those obtained using the three Doppler notch configurations. Therefore, only representative PDPs are presented in Figs. 4 and 5, corresponding to first-order FIR filtering and Doppler notch filtering with a half-width of 0.9 Hz, respectively.

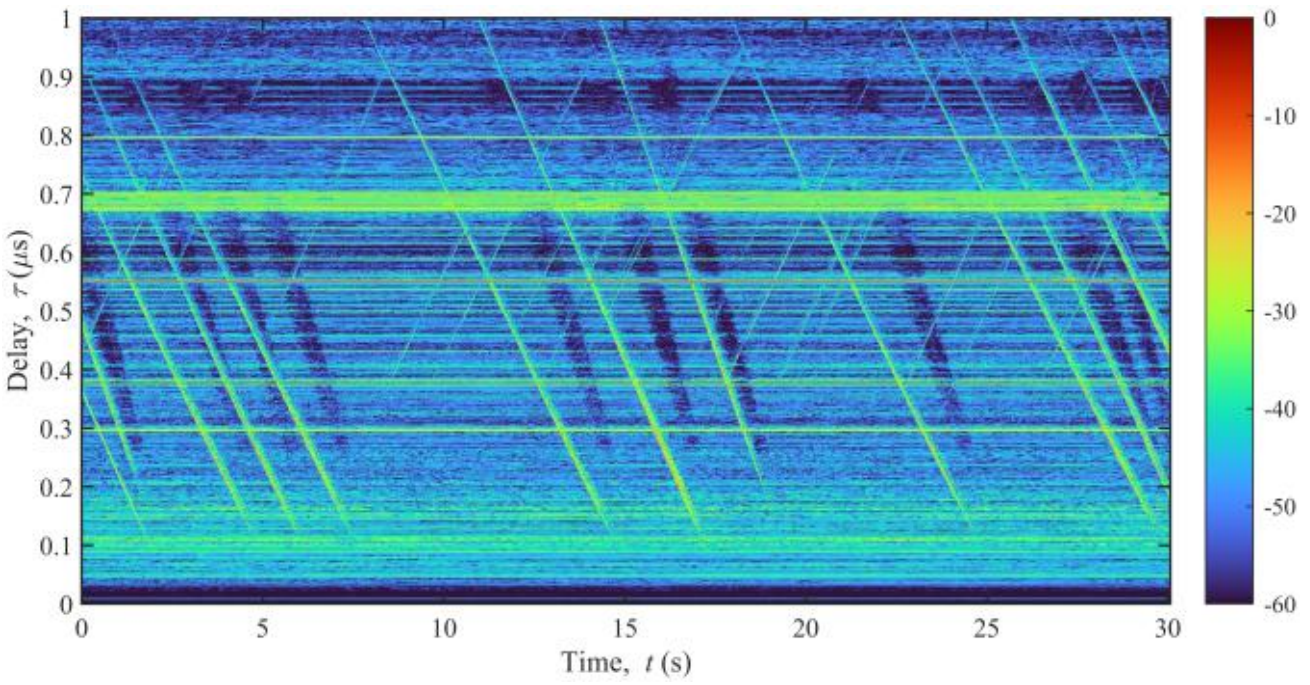

Fig. 3. PDP obtained from the unfiltered CIR.

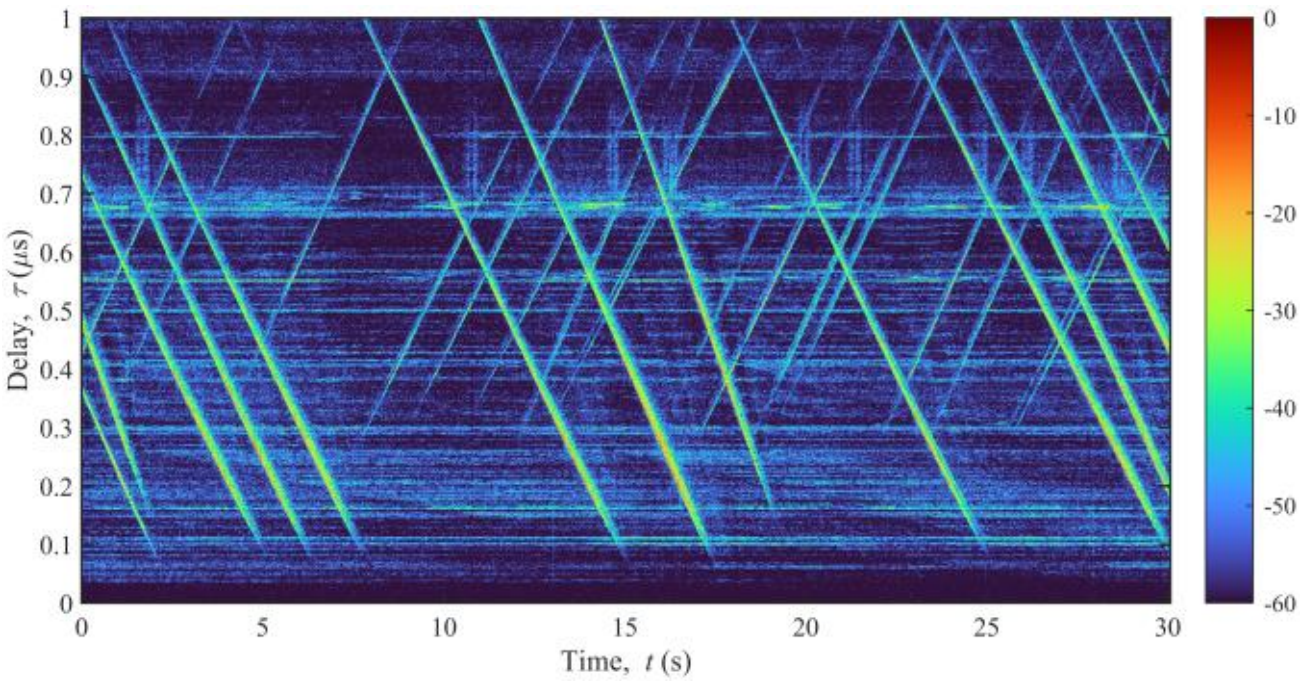

Fig. 4. PDP after first-order temporal FIR filtering.

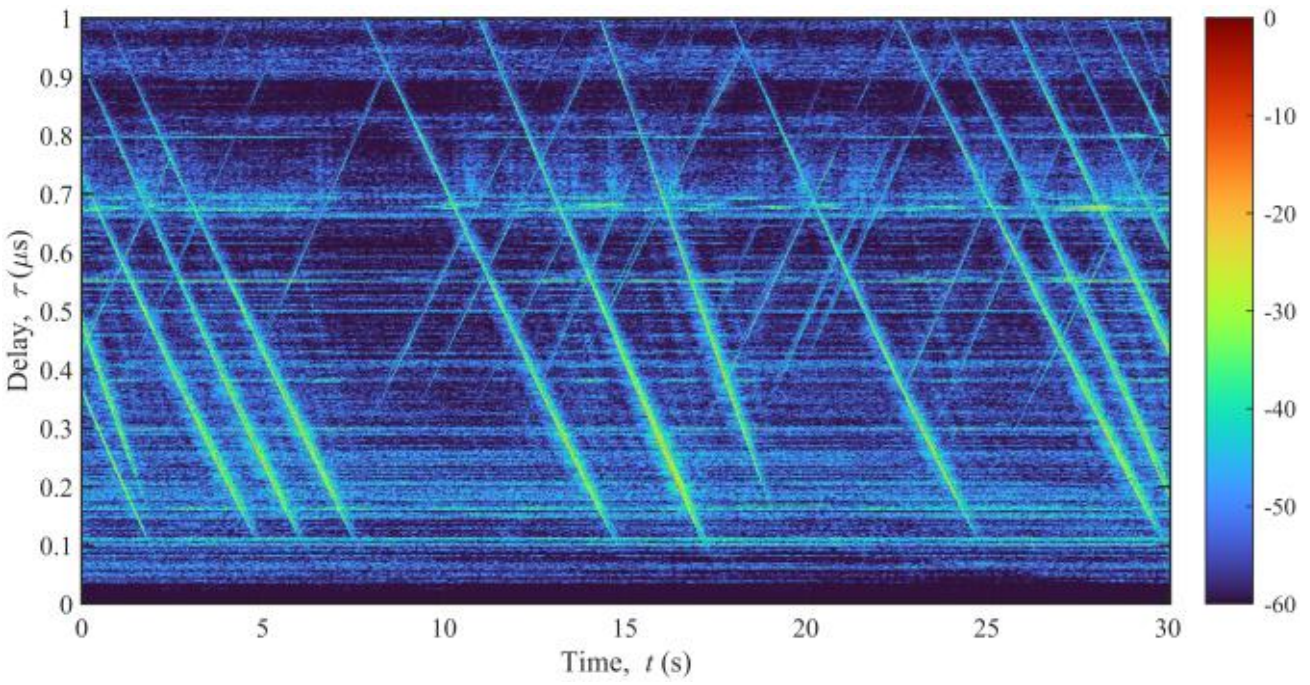

Fig. 5. PDP after Doppler notch filtering, $|f_D| \leq 0.9$Hz.

Both approaches improve the visibility of vehicle-related trajectories, although differences in their displayed intensities can be observed. In particular, some inclined traces appear stronger after FIR filtering, especially those associated with receding vehicles. This behavior is consistent with the Doppler-dependent amplitude responses of the FIR filters, which modify the powers of dynamic multipath components. In contrast, Doppler notch filtering preserves spectral components outside the rejected frequency interval. Increasing the notch half-width from 0.9 to 1.9 Hz produces relatively minor changes in the PDP structure, although wider notches may also suppress slowly varying components associated with moving objects.

To quantitatively evaluate the effects of filtering, the RMS delay spread was calculated for the original and filtered PDPs using a relative power threshold of −30 dB with respect to the maximum of each individual PDP.

Fig. 6 presents the temporal evolution of RMS delay spread over the 30 s observation interval. The unfiltered PDP generally yields lower values than the filtered profiles, suggesting that strong stationary contributions concentrate a substantial fraction of the received power and reduce the apparent influence of more widely distributed reflections from moving vehicles. Consequently, the unfiltered RMS delay spread characterizes the combined stationary and dynamic propagation response rather than predominantly vehicle-related multipath components.

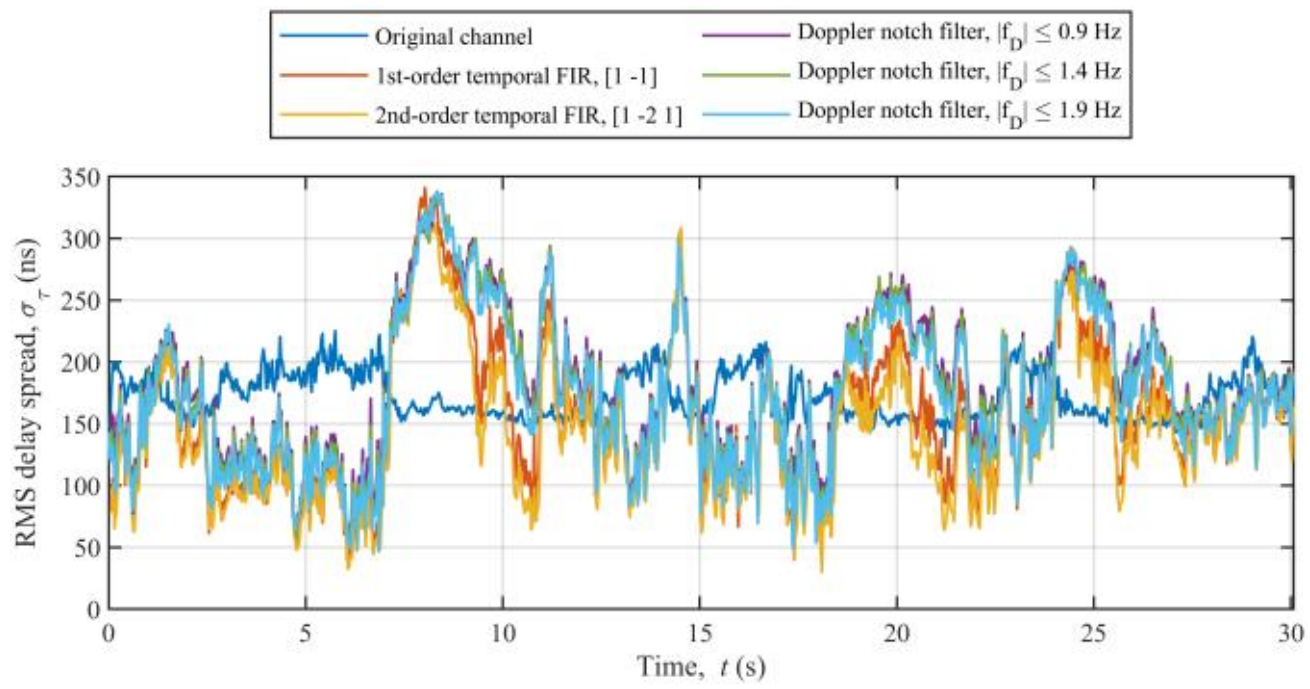

Fig. 6. RMS delay spread versus time for the original channel and after static multipath suppression.

The filtered RMS delay spread trajectories exhibit broadly similar temporal trends, with more pronounced differences during approximately 8–11 s, 18–22 s, and 24–27 s. These discrepancies are consistent with the distinct frequency responses of the applied filters, which modify the relative powers of dynamic multipath components and, consequently, the resulting delay statistics.

Fig. 7 compares the empirical cumulative distribution functions (CDFs) of RMS delay spread, while Table I summarizes the corresponding 10th (P10), 50th (P50), and 90th (P90) percentiles. The unfiltered channel exhibits a relatively narrow distribution, with P10 = 151.7 ns, P50 = 165.0 ns, and P90 = 198.2 ns. In contrast, all filtering methods produce wider distributions, indicating increased variability in the estimated temporal dispersion after suppressing persistent multipath components.

Temporal FIR filtering reduces the median RMS delay spread to 147.6 ns and 137.6 ns for the first- and second-order filters, respectively. Doppler notch filtering yields higher median values, ranging from 165.5 to 173.2 ns, depending on the notch half-width. The differences are more pronounced at the distribution tails. In particular, the P10 values decrease to 79.3–107.4 ns after filtering, whereas P90 increases to 223.6–266.1 ns. Consequently, the interdecile range (P90–P10) increases from 46.5 ns for the unfiltered channel to 144.3–160.1 ns for the filtered cases.

filters, respectively. Doppler notch filtering yields higher median values, ranging from 165.5 to 173.2 ns, depending on the notch half-width. The differences are more pronounced at the distribution tails. In particular, the P10 values decrease to 79.3–107.4 ns after filtering, whereas P90 increases to 223.6–266.1 ns. Consequently, the interdecile range (P90–P10) increases from 46.5 ns for the unfiltered channel to 144.3–160.1 ns for the filtered cases.

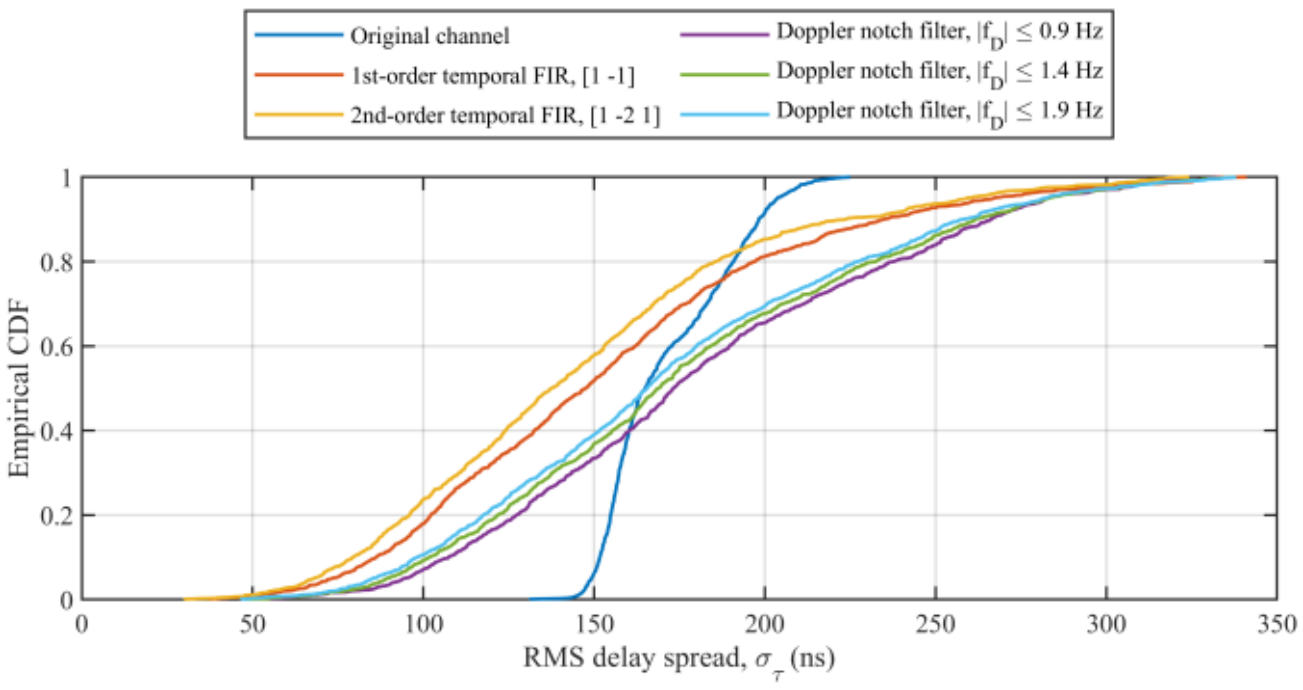

Fig. 7. Empirical CDF of the RMS delay spread for the original channel and after static multipath suppression.

TABLE I. RMS DELAY SPREAD PERCENTILES FOR ORIGINAL AND FILTERED CHANNEL RESPONSES

| **PDP** | **RMS delay spread, $\sigma_\tau$ (ns)** | | |
|---|---|---|---|
| | ***P10*** | ***P50*** | ***P90*** |
| Original channel | 151.7 | 165.0 | 198.2 |
| 1st-order temporal FIR, [1 -1] | 86.1 | 147.6 | 234.8 |
| 2nd-order temporal FIR, [1 -2 1] | 79.3 | 137.6 | 223.6 |
| Doppler notch filter, $\|f_D\| \leq 0.9$ Hz | 107.4 | 173.2 | 266.1 |
| Doppler notch filter, $\|f_D\| \leq 1.4$ Hz | 101.8 | 168.8 | 261.7 |
| Doppler notch filter, $\|f_D\| \leq 1.9$ Hz | 98.1 | 165.5 | 257.9 |

These results demonstrate that static multipath suppression substantially affects the statistical characteristics of RMS delay spread. Although the investigated methods produce similar PDP structures and broadly consistent temporal trends, their distinct frequency responses lead to different distributions of the estimated delay spread. Therefore, the choice of filtering method should be considered when interpreting temporal dispersion parameters derived from time-varying channel measurements.

## V. CONCLUSION

This paper investigated the influence of static multipath suppression on temporal dispersion characterization of a time-varying radio channel at 28 GHz. Two temporal differencing filters and three Doppler-domain notch filtering configurations were evaluated using wideband channel measurements conducted in a dynamic road environment. All investigated methods effectively attenuated persistent multipath components, improving the visibility of reflections associated with moving vehicles.

The results demonstrated that static multipath contributions substantially affect RMS delay spread estimates, with the unfiltered power delay profile exhibiting a narrower range of values than the filtered profiles. Although the investigated filtering methods produced broadly similar dynamic multipath structures, differences in their frequency responses influenced the relative powers of the remaining components and the resulting RMS delay spread distributions. These findings highlight the importance of static multipath suppression when characterizing propagation effects associated with moving scatterers, while emphasizing that the selected filtering method must be considered when interpreting temporal dispersion estimates.

## ACKNOWLEDGMENT

This work was co-funded by the National Science Centre, Poland, under the OPUS call in the Weave programme under research project no. 2021/43/I/ST7/03294 acronym 'MubaMilWave'; by the Czech Science Foundation under project no. 23-04304L; by the Military University of Technology under project no. UGB 531-000128-W300-22/2026 "Analysis of the possibility of using modern technologies in communication systems and electronic warfare"; and by the MeitY, Government of India, through the Chips-to-Startup (C2S) program under project no. EE-9/2/2021-R&D-E.